\documentclass[12pt]{article}

\def\gtap{\raisebox{-.4ex}{\rlap{$\sim$}} \raisebox{.4ex}{$>$}}

\def\ra {\rightarrow}
\def\qbq {\overline qq}
\def\qbqg {\overline qqg}

\def\qbqbqq {\overline{qq}qq}
\def\ubu {\overline uu}
\def\dbd {\overline dd}

\def\kbk {\overline KK}
\def\pp {\pi\pi}
\def\qbqg {\overline qqg}
\def\sbs {\overline ss}
\def\jp {\Psi}
\def\journal{\topmargin 0.75in   \oddsidemargin 0in
        \headheight 0pt \headsep 0pt
        \textwidth 6.5in 
\textheight 9in 
        \marginparwidth 1.5in
        \parindent 2em
        \parskip .5ex plus .1ex         \jot = 1.5ex}
\journal

\begin{document}
\begin{titlepage}

\noindent September 18, 2006      \hfill    LBNL-61631\\

\begin{center}

\vskip .5in

{\large \bf Hunting the Scalar Glueball: Prospects for BES III}

\vskip .5in

Michael S. Chanowitz\footnote{Email: chanowitz@lbl.gov}

\vskip .2in

{\em Theoretical Physics Group
     Ernest Orlando Lawrence Berkeley National Laboratory\\
     University of California\\
     Berkeley, California 94720}
\end{center}

\vskip .25in

\begin{center}
Presented at 
CHARM 2006,\\ International Workshop on Tau-Charm Physics, \\
June 5 - 7, 2006, Beijing, China\\
To be published in the proceedings. 
\end{center}

\vskip .25in

\begin{abstract}

The search for the ground state scalar glueball $G_0$ is reviewed.
 Spin zero glueballs will have unique dynamical properties if the
 $<G_0|\overline qq>$ amplitude is suppressed by chiral symmetry, as
 it is to all orders in perturbation theory: for instance, mixing of
 $G_0$ with $\overline qq$ mesons would be suppressed, radiative $\jp$
 decay would be a filter for new physics in the spin zero channel, and
 the decay $G_0 \rightarrow \overline KK$ could be enhanced relative
 to $G_0 \rightarrow \pi \pi$. These properties are consistent with
 the identification of $f_0(1710)$ as the largely unmixed ground state
 scalar glueball, while recent BES data implies that $f_0(1500)$ does
 not contain the dominant glueball admixture. Three hypotheses are
 discussed: that $G_0$ is 1) predominantly $f_0(1500)$ or 2)
 predominantly $f_0(1710)$ or 3) is strongly mixed between $f_0(1500)$
 and $f_0(1710)$.

\end{abstract}

\end{titlepage}
\newpage

\section{Introduction}	

Glueballs are a dramatic consequence of the local, unbroken,
non-Abelian symmetry that is the unique defining property of QCD. 
Non-Abelian gauge bosons (gluons) carry the non-Abelian charge and
therefore interact directly with one another. Because the symmetry is
unbroken, charge is confined and singlet combinations of two or more
gauge bosons form bound states. In QED the Abelian gauge boson, the
photon, has no electric charge, the force does not confine, 
and there is no ``lightball'' counterpart of the QCD
glueball.

The prediction that glueballs exist is simple and fundamental but has
proven difficult to verify. We expect their discovery soon, for two
reasons. First, BES III will provide huge $J/\Psi$ data samples ---
potentially several billion --- allowing definitive studies of $J/\Psi$
decay and, especially, partial wave analysis of the glueball-preferred
radiative $J/\Psi$ decay channel. Second, in roughly the same time frame,
lattice QCD (LQCD) will provide reliable unquenched predictions for
the glueball spectrum, mixing, and decays. This powerful combination
of theory and experiment should suffice to finally resolve this
fundamental and difficult problem.

Quenched LQCD calculations have verified the naive expectation that
glueballs exist. The most recent quenched results\cite{chenetal} put
the ground state scalar mass at $m_G = 1710 \pm 50 \pm 80$ MeV.
Glueballs are hard to identify because they are not easily
distinguished from ordinary $\overline qq$ mesons with which they can
mix, and because dynamical properties, such as decay widths and
branching ratios, are not understood. The problem is further
complicated by the likely presence of $\qbqg$ hybrids and possibly
also $\qbqbqq$ states, with which they may also be confused and mix.

For now we rely on a few simple ideas: 
\begin{itemize}
\item Glueballs are extra states, beyond the $\overline qq$
spectrum. To exploit this we must understand the ``ordinary''
$\overline qq$ spectrum very well, using data from 
$\jp$, $B$, and $Z$ decays, and from 
$\overline pp$, $\pi p$, $\gamma \gamma$, and $\gamma N$
scattering. It is already clear that there are indeed ``extra''
$I,J^{PC}= 00^{++}$ states in the mass region where the scalar
glueball is expected.

\item Glueballs couple strongly to gluons so they are prominent in
radiative $\jp$ decay, which proceeds via $\Psi \rightarrow \gamma
gg$. They couple weakly to photons so they are not prominent in
photon-photon scattering.\footnote { While neither generic nor
physically motivated, it is possible to arrange singlet-octet flavor
mixing so that a $\qbq$ meson also has a small or vanishing $\gamma
\gamma$ coupling.  }

\item Glueballs are flavor singlets so their decays should be
$SU(3)_F$ symmetric. However, this may not be true of spin zero
glueball decays because of chiral suppression, as discussed below.

\end{itemize}

Since hybrid and $\qbqbqq$ states are also ``extra'' states and some
are flavor singlets, the only distinguishing property unique to
glueballs is their strong coupling to the color singlet digluon
channel.  Radiative $\jp$ decay then plays a very special role because
for heavy quarks $Q$ we know reliably from perturbation theory that
the leading mechanism is $\Gamma(\Psi(\overline QQ) \ra \gamma X)
\simeq \Gamma(\Psi(\overline QQ) \ra \gamma gg)$, with\cite{mc-ov}
$$ 
{{\Gamma(\Psi \ra \gamma gg)} \over {\Gamma(\Psi \ra ggg)}} = 
         {{16\alpha} \over {5\alpha_S}} \simeq 0.09.        
\eqno{(1)}
$$ Using $B(\Psi \ra ggg) \simeq B(\Psi \ra {\rm hadrons})_{\rm
direct} = 0.71$, we obtain $B(\Psi \ra \gamma X) \simeq 0.06$. This is
consistent with the only attempted inclusive measurement,\cite{mk2}
verifying that the perturbative mechanism is at least roughly
correct.\footnote{Although the shape is distorted by resonances, the
measured rate for photons with $\geq 60\%$ of the beam energy is
consistent with QCD,\cite{mk2} as expected for ``global duality.''}
The leading partial waves of the digluon in perturbation theory are
$J^{PC}= 0^{++}, 0^{-+}, 2^{++}$,\cite{billoireetal} corresponding
precisely to the quantum numbers of the lightest glueballs. Radiative
$\jp$ decay is then a copious source of photon-tagged, color-singlet
gluon pairs, perfectly matched to the expected masses and quantum
numbers of the lightest glueballs.

Heavier quarkonia cannot compete: e.g., for equal luminosity, the
number of events in the glueball mass region for $\Upsilon$ radiative
decay is smaller by a factor $\simeq 10^2 \times 4 \times 10 = 4000$,
where $10^2$ reflects the observed peak cross sections, 4 is from the
square of the quark charges, and the final 10 is from the branching ratio
into the relevant digluon mass region.  Radiative $\jp$ decay is the
ideal glueball hunting ground, for which BEPC II/BES III will be the
premier world facility.

Glueballs are sticky because they couple strongly to gluons and 
weakly to photons. The stickiness of particle $X$ is defined 
as\cite{sticky}
$$
S_X = {{\Gamma(\Psi \ra \gamma X)} \over {\Gamma(X\ra \gamma \gamma)}}
  \times {{PS(X\ra \gamma \gamma)} \over {PS(\Psi \ra \gamma X)}}, 
\eqno{(2)}
$$ 
where $PS$ denotes phase space. We consider
stickiness ratios, since glueballs will typically be much stickier
than $\qbq$ mesons, $\qbqg$ hybrids, or $\qbqbqq$ states. It is worth
considering if the high luminosity at BEPC will make it feasible to study
$\gamma \gamma$ scattering at BES III despite the low beam energy.

\section{Chiral Suppression}

If chiral symmetry breaking in glueball decay is dominated by quark
masses, then the coupling of a spin zero glueball to light $\qbq$
pairs is chirally suppressed,\cite{msc-cs}
$$
<G_0|\qbq> \propto m_q/m_G,             \eqno{(3)}
$$ 
like the suppression of $\pi \to e \nu$, though different in
detail. This is easily understood: for $m_q=0$ chiral symmetry
requires the quark and antiquark to have equal chirality, hence
unequal helicity, implying nonvanishing net angular momentum, so that
the amplitude must vanish for $J=0$ in the chiral limit.  Chiral
suppression, eq. (3), is valid to all orders in perturbation
theory.\cite{msc-cs,chaoetal} Explicitly at leading order\cite{msc-cs}
$$ 
{\cal M}(G_0 \ra
\overline qq) = -f_0\alpha_S\ {16\pi\sqrt{2} \over 3}
                      \ {m_q\over \beta} \ {\rm
log} {1 + \beta \over 1 - \beta} \ \overline u_3 v_4 \delta_{ij}.
\eqno{(4)}
$$ 
where $f_0$ is the effective $G_0gg$ coupling and $\beta$ is the 
quark velocity in the $G_0$ cms. 

However, there is no limit in which (4) is a
reliable estimate of the magnitude. Even for
$m_G \ra \infty$ the $t$ and $u$ channel quark exchange amplitudes are
not under perturbative control.  We cannot calculate the magnitude of
the amplitude but we know it is suppressed of order $m_q/m_G$ to
all orders in perturbation theory.

Nonperturbative chiral symmetry breaking might lift the chiral
suppression, as suggested\cite{zj} in the context of the liquid
instanton model. A reliable, model-independent, nonperturbative method
is needed to decide: for now LQCD is the only game in town.  Early
results are equivocal, as discussed below.  The phenomenological
proposal that chiral symmetry is restored\cite{lg1} in the baryon and
meson spectra for $\gtap{\rm O}(2)$ GeV suggests that nonperturbative
chiral symmetry breaking is not large at the glueball mass scale, and
in fact motivated the suggestion that a ``high lying'' scalar
glueball would not mix strongly with $\ubu + \dbd$ mesons.\cite{lg2}

Chiral suppression has important consequences for spin zero
glueballs. Mixing with light ($u,d,s$) mesons is
suppressed of order ${\rm{O}(m_q/m_G)}$, so that $J=0$ glueballs are more
likely than $J\neq 0$ to be largely unmixed. (Although mixing
amplitudes are suppressed, mixing angles can be large if the
quenched glueball and meson states happen to be extremely
degenerate.) To the extent $G_0-M_0(\qbq)$ mixing does occur, it should be
dominated by $\sbs$ components. Mixing with hybrids and four-quark
states is not suppressed. 

A second consequence is that radiative $\jp$
decay becomes a filter for new physics in the $J=0$ channel, since at
leading order the exclusive amplitude $\jp \to \gamma X$ is
proportional to $<gg|X>$, so that radiative decays to spin zero light
quark mesons, $X=M_0(\qbq)$, are suppressed, and, to the extent they
do occur, favor $J=0$ strangeonium, $M_0(\sbs)$, over $M_0(\ubu
+\dbd)$. Radiative decays to $J=0$ glueballs, hybrid, and four quark
states are not suppressed.

A third consequence is that $\qbq$ decays of $J=0$ glueballs favor the
heaviest quark $q$. If the multibody decays have discernible jet
structure, decays to two jets will contain leading strange particles
if $m_G < 2m_D$ or two charm particles if $2m_D < m_G < 2m_B$, while
the leading particles of three jet decays are flavor symmetric. For
$m_G < 2m_D$ we could then see an increase in leading strange
particles in events with high thrust.  For $m_G \simeq 1700$ MeV the
partonic preference for $\sbs$ over $\ubu + \dbd$ decays would favor
$\kbk$ favored over $\pp$ if hadronization of $G_0 \to \qbq$ is an
important short distance mechanism for two meson decays at this mass
scale. Another possibility is that the dominant short distance
mechanism is $G_0 \to \overline{qq}qq$, which would imply that $\kbk$
and $\pp$ are more nearly equal.\cite{chaoetal} Or both mechanisms
could be important and the ratio could lie between the two
predictions.

The existing evidence from LQCD is preliminary and equivocal. A
quenched study\cite{svw} of scalar glueball decay to two pseudoscalar
mesons found that the amplitude decreases with the meson mass $m_P$ at
a rate consistent with the $m_P^2$ dependence expected for chiral
suppression (since $m_q \propto m_P^2$).  A study of
$G_0-\sbs$ mixing\cite{lw99} found a small mixing energy, $E_M = 43
\pm 31$ MeV, also as expected, but was not consistent with $E_M \propto
m_q$. Given the small $\sim 1\sigma$ ``signal,'' this calculation may
have lacked the precision needed to obtain the $m_q$
dependence. Another study found large mixing at the strange quark mass
but the lattice granularity was far from the continuum limit.\cite{mm}
All these studies extrapolated from quark masses near or above the
strange quark mass and did not directly probe the region of the up and
down quark masses. They should be revisited with today's computing
power, to simultaneously explore the chiral and continuum limits. A
quenched calculation of mixing would suffice to determine whether
chiral suppression occurs or not.

\section{Experimental Status of the Scalars}

The most recent quenched LQCD calculation obtained $m_G= 1710 \pm 50
\pm 80$ MeV for the scalar glueball mass.\cite{chenetal}
Experimentally there are too many isoscalar, scalar mesons between 1.4
and 2 GeV to be explained by the naive quark model alone. I assume
$f_0(600)$ and $f_0(980)$ are cryptoexotic $\qbqbqq$ states.\cite{rlj}
The p-wave $\qbq$ scalar nonet is likely to lie in the region of the
other spin-triplet p-wave nonets, with isoscalars roughly between
$\sim 1250$ and $\sim 1600$ MeV.  Between $\sim 1400$ and 2000 MeV
there are five $I,J^{PC}= 0,0^{++}$ states: $f_0(1370)$, $f_0(1500)$,
and $f_0(1710)$ are well known, while $f_0(1790)$ and $f_0(1810)$ were
recently discovered by BESII.  It seems likely that some of these five
states have gluon constituents: we have probably seen the scalar
glueball although we cannot yet identify it. I will briefly discuss
the possibility that the scalar glueball is predominantly 1)
$f_0(1500)$,\cite{a-c} or 2) $f_0(1710)$\cite{svw,lw99,msc-cs}, or 3)
is shared by both in a maximally mixed glueball-strangeonium duo.

\subsection{$f_0(1500)$ is the glueball}

Since $f_0(1500)$ is produced in ``gluon rich'' $\overline pp$
annihilation and $\pi N$ central production while $f_0(1710)$ decays
prominently to $\kbk$, it was natural to consider the hypothesis that
$f_0(1500)$ is the scalar glueball and $f_0(1710)$ is the $\sbs$
scalar nonet partner of $f_0(1370)$.\cite{a-c} However, the
dynamics of $\overline pp$ annihilation and $\pi N$
central production are not as well understood as radiative $\jp$
decay, which we know from perturbation theory is a copious source of
color singlet gluon pairs in the relevant mass region. It is then
problematic that $f_0(1500)$ is not strongly produced in radiative
$\jp$ decay, in old Mark III data\cite{mk3} and more recently in BES
II data,\cite{gpp} while the $f_0(1710)$ is prominent in both data
sets.\cite{mk3,1710-bes2} In addition, quenched LQCD calculations find
that the glueball is $\simeq 200$ MeV lighter than scalar
strangeonium,\cite{lw99,mm} while the opposite ordering is required by
this hypothesis.\cite{c-z}

The recently reported BES II partial wave analysis of $\jp \to \gamma
\pp$ is an important result.\cite{gpp} The rates for $\pi^+ \pi^-$ and
$\pi^0 \pi^0$ agree, a critical check since $\gamma \pi^0 \pi^0$ is
free of the large $\jp \to \pi^0 \rho^0$ background that afflicts
$\gamma \pi^+ \pi^-$. This is the best channel to search for $\jp \to
\gamma f_0(1500)$ because of the simplicity of the two pion final
state and because $B(f_0(1500) \to \pi\pi) = 0.349 \pm 0.0223$ is
large.\cite{pdg} BES II finds only a small possible signal, $B(\jp \to
\gamma f_0(1500)) \times B(f_0(1500) \to \pi^+ \pi^-) = (6.7 \pm 2.8)
\cdot 10^{-5}$, implying
$$
B(\jp \to \gamma f_0(1500)) = (2.9 \pm 1.2)\cdot 10^{-4}.   \eqno{(5)}
$$ 
This is small, viewed
either as a fraction of all radiative decays or compared to the 
rate for $f_0(1710)$, for which the lower limit is six times larger,
$$
B(\jp \to \gamma f_0(1710)) \geq (16.2 +3.0 -2.4)\cdot 10^{-4},  \eqno{(6)}
$$ from just the $\kbk$ and $\eta \eta$ decay modes, using $B(\jp \to \gamma
f_0(1710)) \times B(f_0(1710) \to \kbk) = (11.1 +1.7 -1.2) \cdot
10^{-4}$ from the BES II bin-by-bin fit\cite{1710-bes2} and
$B(f_0(1710) \to \eta \eta)/B(f_0(1710) \to \kbk) = 0.48 \pm
0.15$.\cite{wa102,pdg} The $f_0(1710)$ probably has other decays,
especially multibody modes which are difficult to measure, so the
inclusive rate for $\jp \to \gamma f_0(1710)$ is likely to be
appreciably larger.\footnote{The lower limit (6) would increase to
$20.2\cdot 10^{-4}$ if the BES II result for $\jp \to \gamma f_0(1710)
\to \gamma \pp$ were included. I have not included it here because 
a smaller value for the $\pp$ mode is implied by the BES II 95\% upper 
limit from $\jp \to \omega \pp$ as reviewed in Section 3.2.}  The
additional statistical power of BES III may be needed to analyze the
multibody modes.

An early attempt to study the four pion channel, $\jp \to \gamma +
4\pi$, was made with the 8M event BES I data sample.\cite{4pi-bes1}
Given the complexity of the analysis, including assignment of the four
pions to two-isobar intermediates, this would be challenging even with
the 58M BES II data set. The BES I results from $f_0(1500) \to 4\pi$
are not consistent with the BES II $f_0(1500) \to 2\pi$ results. From
BES I, $B(\jp \to \gamma +f_0(1500)) \times B( f_0(1500) \to \pi^+
\pi^- \pi^+ \pi^-) = (3.1 \pm 0.2 \pm 1.1) \cdot 10^{-4}$, which
implies $B(\jp \to \gamma +f_0(1500)) \times B( f_0(1500) \to 4\pi) =
(7.0 \pm 2.5) \cdot 10^{-4}$ using isospin (with the decay chain $f_0
\to \sigma \sigma \to 4\pi$ used in the BES I analysis) to include
neutral pion modes. Using $B(f_0(1500) \to 4\pi) = 0.495 \pm
0.033$,\cite{pdg} the BES I measurement implies the inclusive rate
$B(\jp \to \gamma +f_0(1500)) = (14 \pm 5.1) \cdot 10^{-4}$, a factor
5 larger (and $2\sigma$ higher) than eq. (5) from the BES II
$f_0(1500) \to 2\pi$ measurement. The $\pp$ measurement must be given
greater weight, since it considers a much simpler final state, uses
the upgraded BES II detector, and is based on seven times more
statistics.

Another problem is posed by hadronic $\jp$ decay data. The $f_0(1710)$
is produced prominently in $\jp \to \omega f_0(1710) \to \omega
\kbk$\cite{bes2-omegakk} but not in $\jp \to \phi f_0(1710) \to \phi
\kbk$,\cite{bes2-phikk} contrary to the OZI rule if $f_0(1710)$ is an
$\sbs$ state. But the OZI rule does correctly describe the pattern of
the four decays to the ideally mixed tensor mesons, $\jp \to
\omega/\phi + f_2(1270)/f_2(1525)$.\cite{pdg} If $f_0(1710) = \sbs$,
it is necessary to assume that dynamics in the $J=0$ channel somehow
makes the doubly OZI suppressed rate not just comparable to the
singly suppressed one but $\sim 5$ times larger (see Close and
Zhao\cite{c-z}).

\subsection{$f_0(1710)$ is the glueball}

Another possibility is that $f_0(1710)$ is the scalar
glueball.\cite{svw,lw99,msc-cs} This is consistent with its prominence
in radiative $\jp$ decay, eq. (6), and its mass, in the middle of the
range of the most recent quenched LQCD prediction.\cite{chenetal}
Chiral suppression could then explain the absence of strong
glueball-meson mixing.  It is also clearly seen in $\jp \to \omega
f_0(1710) \to \omega \kbk$ with virtually the same mass and width as
in $\jp \to \gamma f_0(1710) \to \gamma \kbk$ but it is not seen in
$\jp \to \omega f_0(1710) \to \omega \pp$ despite the much greater
statistics of the $\omega \pp$ channel, yielding a robust 95\% CL
upper limit, $B(f_0(1710) \to \pp)/B(f_0(1710) \to \kbk) <
0.11$.\cite{bes2-omegakk} However, a possible indication of $f_0(1710)
\to \pp$ appears in $\jp \to \gamma \pp$ where BES II finds a scalar
at $1765 ^{+4}_{-3}\pm 12$ MeV with $\Gamma = 145 \pm 8 \pm 69$ MeV.
If attributed to $f_0(1710)$ it implies $B(f_0(1710) \to
\pp)/B(f_0(1710) \to \kbk) = 0.41^{+0.11}_{-0.17}$, which is
1.8$\sigma$ above the 95\% upper limit from $\jp \to \omega +
\pp/\kbk$. The signal at 1765 could also be due to the $f_0(1790)$
seen in $\jp \to \phi \pp$\cite{bes2-phikk} or it could be the result of
interference between $f_0(1710)$ and $f_0(1790)$. If the
stronger upper limit from $\jp \to \omega + \pp/\kbk$ prevails, chiral
suppression could explain the suppression of the $\pp$
mode,\cite{msc-cs} which is also consistent with a quenched LQCD
calculation.\cite{svw} The problem then would be to find the
strangeonium component of the scalar $\qbq$ nonet, since neither
$f_0(1370)$ nor $f_0(1500)$ seem to have much $\sbs$ content.

A possible solution is suggested by data from charmless $B$ meson
decays, $B \to K \kbk$ and $B \to K \pp$.\footnote {I thank Alex
Bondar for telling me of these results.} Belle\cite{belle} and
BABAR\cite{babar} both see a strong signal for a scalar meson near
1500 MeV which decays to $\kbk$ but not to $\pp$, although the
amplitude analysis of the $\kbk$ channel has significant model
dependent ambiguities requiring further study. This object cannot be
the previously observed $f_0(1500)$, for which $B(\kbk)/B(\pp)=0.241
\pm 0.028$,\cite{pdg} but it could be the missing $\sbs$ scalar.  If
it is the $\sbs$ scalar and $f_0(1370)$ is its $\ubu + \dbd$ nonet
partner, then an explanation is needed for the previously observed
$f_0(1500)$.

\subsection{$f_0(1500)$ and $f_0(1710)$ share the glue}

Since the amplitude ${\cal M}(gg \to \qbq)_{J=0} \propto m_q$ is
dominated by forward and backward scattering, the effective running
mass $m_q$ must be evaluated at a low energy scale of order
$\Lambda_{\rm QCD}$. The effective masses are larger than their
``current quark'' values but the hierarchy $m_u,m_d \ll m_s$ is
maintained. The effective value of $m_s$ might then be large enough
that ${\cal M}(gg \to \sbs)_{J=0}$ is not chirally suppressed while
${\cal M}(gg \to \ubu + \dbd)_{J=0}$ is. The scalar glueball could
then mix strongly with strangeonium but not with $\ubu +\dbd$
mesons. The quenched scalar glueball, $G_0(gg)$, and the strangeonium
meson, $f_0(\sbs)$, which are expected to have masses near one
another, could then mix maximally, yielding the eigenstates
$$
f_0^{\pm} \simeq {1 \over \sqrt{2}} [G_0(gg) \pm f_0(\sbs)].    
                     \eqno{(7)}
$$

Now consider the amplitudes $<gg|f_0^{\pm}>$ and $<\sbs|f_0^{\pm}>$.
The first determines the rate for $\jp \to \gamma f_0^{\pm}$, while I
will assume the latter is the dominant partonic mechanism for
$f_0^{\pm} \to \kbk$. We choose the phases of the wave functions so
that the ``elastic'' amplitudes, $<gg|G_0(gg)>$ and
$<\sbs|f_0(\sbs)>$, are real and positive. If the ``inelastic''
amplitudes, $<\sbs|G_0(gg)>$ and $<gg|f_0(\sbs)>$, have equal phase
and that phase is real relative to the ``elastic'' amplitudes, the
rates for $\jp \to \gamma f_0^{\pm}$ and $f_0^{\pm} \to \kbk$ would
replicate the experimentally observed pattern. One state, say $f_0^+
\simeq f_0(1710)$, would be produced prominently in radiative $\jp$
decay and would decay prominently to $\kbk$, because of constructive
interference of the $gg$ and $\sbs$ components, while the
corresponding $f_0^- \simeq f_0(1500)$ amplitudes would be suppressed
by destructive interference. This modified chiral suppression scenario
could be tested in quenched LQCD studies of mixing between the
quenched scalar glueball and the $f_0(\ubu + \dbd)$ and $f_0(\sbs)$
scalar mesons.\footnote {In this scenario, radiative $\jp$ decay
filters out $J=0$ $\ubu + \dbd$ mesons but not $\sbs$.} In this
connection it is amusing that chiral symmetry restoration is seen
clearly in the spectrum of $u,d$-quark baryons but not in the strange
baryon spectrum.\cite{jps}

\section{Discussion}

BES III at BEPC II will begin operation in 2007. At design luminosity 
it will accumulate several billion $\jp$ decays in a single year, 
enabling definitive partial wave analysis of the decay products. 
We can look forward to better understanding of the scalar glueball 
candidates, including their multibody decays. During the BEPC II lifetime 
LQCD should begin to contribute reliable unquenched calculations of 
the spectrum, mixing and decays.  In particular, LQCD can determine 
if chiral suppression survives nonperturbative effects and, if so, 
how it effects mixing and decays. The combination of BES III and 
LQCD should allow us to finally identify and study the scalar glueball, 
as well as glueballs and hybrids of other quantum numbers. 

\section*{Acknowledgments}

I wish to thank Alex Bondar, Robert Cahn, and Zoltan Ligeti for
discussions.

\noindent{\small This work was supported in part by the Director, Office of
Science, Office of High Energy and Nuclear Physics, Division of High
Energy Physics, of the U.S. Department of Energy under Contract
DE-AC03-76SF00098.}

%


\end{document}